\documentstyle[11pt,paspconf]{article}
\input epsf

\begin{document}

\title{CBR Polarization Experiments}

\author{S. T. Staggs\altaffilmark{1} and J. O.
Gundersen\altaffilmark{2}}
\affil{Physics Department, Princeton University, Princeton, NJ
08544}

\author{S. E. Church}
\affil{Physics Department, Stanford University, Stanford, CA
94305}

\altaffiltext{1}{Alfred P. Sloan Fellow}
\altaffiltext{2}{Robert H. Dicke Fellow}

\begin{abstract}
        A handful of new experiments aimed at measuring the
as-yet-undetected polarization of the cosmic background radiation
(CBR) are described, with somewhat more detail given for
three experiments with which the authors are associated.
\end{abstract}



\section{Introduction}

\begin{figure}[p]
\epsfxsize=2.75cm \epsfbox[70 520 170 620]{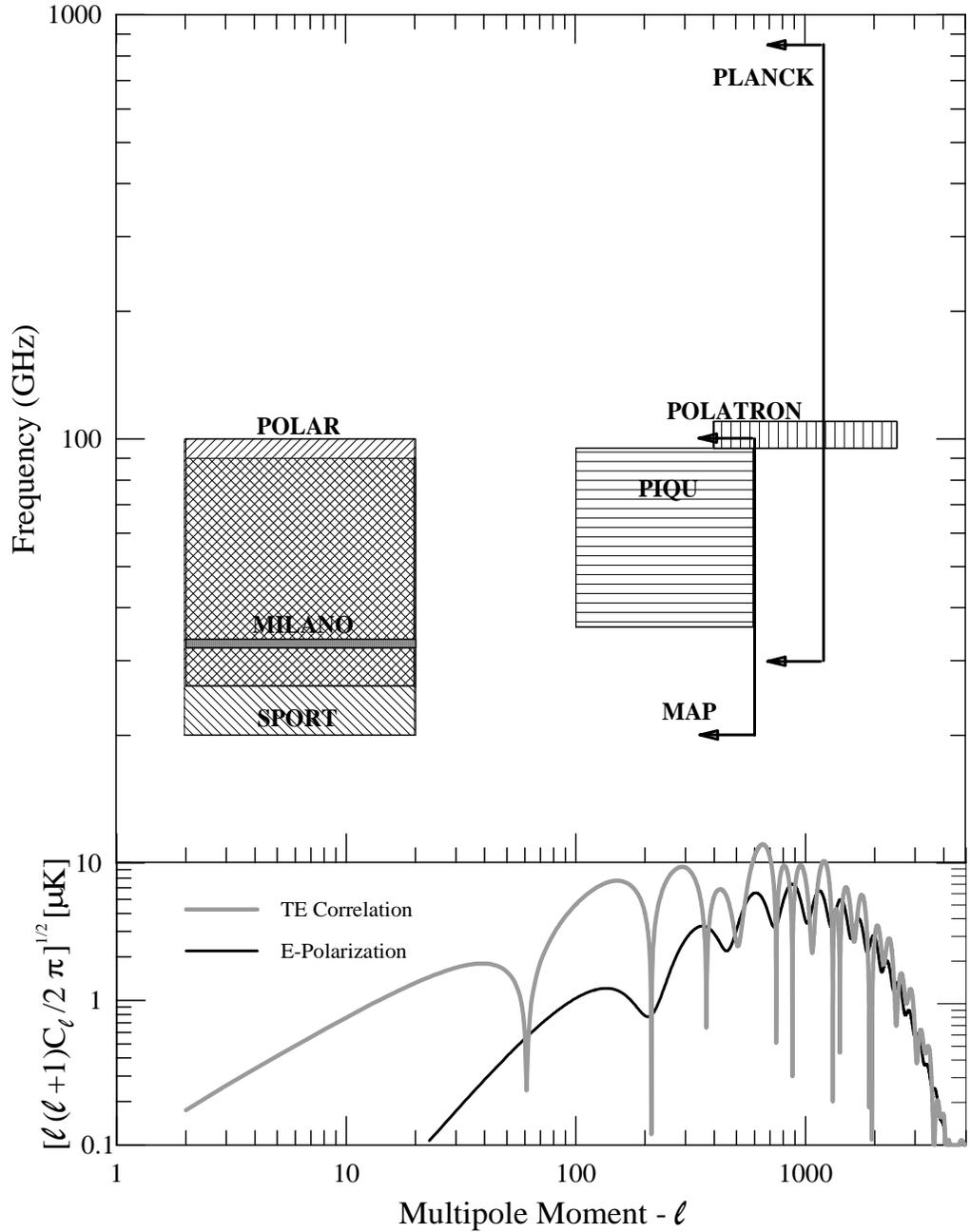}
\vspace{5.0in}
\caption{ A figure schematically indicating the multipole and
frequency coverage of current and upcoming experiments.  The frequency
coverage shown indicates the {\em range} of frequencies probed by
the experiments; in no case is the frequency coverage continuous.
See Table \ref{tbl-1} for more information.
The lower figure shows the E-polarization spectrum and the
absolute value of the TE correlation expected for a typical CDM
model. (The actual parameters are $\Omega_b = 0.05$, $\Lambda =
0$, $h = 0.65$, and $\Omega_{tot}=1$.)  For an explanation of
``E" polarization, see, for example, Zaldarriaga \& Seljak, 1997.
The spectra were generated with CMBFAST (Seljak \& Zaldarriaga
1996).} \label{fig-lvsnu}
\end{figure}

Since Penzias \& Wilson published their groundbreaking report of
the existence of the CBR (1965), experimentalists have been
checking to see if the CBR is polarized. CBR polarization
experiments are challenging.  Predicted  polarization signals are
an order of magnitude below the levels at which CBR anisotropy
experiments have only  just begun to detect signals. Typical
polarization signals are a few $\mu$K.  The most significant
challenges include achieving adequate  statistical sensitivity,
limiting systematic errors sufficiently to detect the small
polarization signal,  and discriminating the CBR polarization
from foreground sources such as galactic synchrotron radiation.
To date, no polarization has been detected, but several
experiments which have just begun operation or plan to within the
year are designed to reach the sensitivities required to detect
polarized signals of about the  size predicted by CDM models.
Figure \ref{fig-lvsnu}  shows the angular resolution and frequency coverage
of
the new and recent experiments.  Some details about the
experiments are listed in Table \ref{tbl-1}.

After a brief inspection of Figure \ref{fig-lvsnu}, one sees that
the experimentalist is faced with a difficult choice.  Should
s/he design an experiment to search for a CBR polarization signal
at large angles, where the spectrum codes information about
reionization and the existence of tensor fluctuations?  Such an
experiment has the potential of enormous scientific payback if
the CBR exhibits the coherent oscillation anisotropy
spectrum predicted by models with adiabatic initial fluctuations
amplified by a period of inflation.  Also, a horn antenna can be
used to attain beamwidths of a few degrees, without the need for
reflectors which might introduce certain systematic errors.
However, the aforementioned models
predict that the CBR polarization anisotropy will be very small,
less than $0.1$~$\mu$K, at angular scales larger than a couple of
degrees.
Extant limits at those angular scales are more than three orders of
magnitude greater than the expected signal.
The other choice for the experimentalist is to design an experiment
to probe the polarization anisotropy at smaller angular scales,
where the predicted signal from standard CDM is larger. These
experiments must contend with polarized offsets due to the use of
reflectors (or lenses).

At present, groups are pursuing both lines of attack, as
indicated in Figure \ref{fig-lvsnu}.    The space-based MAP and Planck
satellites plan to collect data from large fractions of the sky
with small resolutions, which allows probing of small and large
angular scales at once. Ground-based experiments complement the
satellite missions because of their ability to probe specific
regions of sky more deeply. MAP, for example, is expected to only make a
statistical detection of the polarization anisotropy, while
ground-based experiments should be able to detect polarization
directly.  Also, since no detections of polarization have yet
been claimed, experimenters and analysts may not yet have
beaten down all the relevant systematic errors which will plague
the measurements; the ground-based experiments will address
systematic effects as they become evident.  Experience on the
ground may prove helpful to the satellite experiments.
A further distinction is that the
current crop of ground-based experiments
compare the orthogonal linear polarization components
within a single beam at a time rather than comparing them between beams
separated on the sky.

\begin{figure}[p]
\epsfxsize=3.0cm \epsfbox[120 520 220 620]{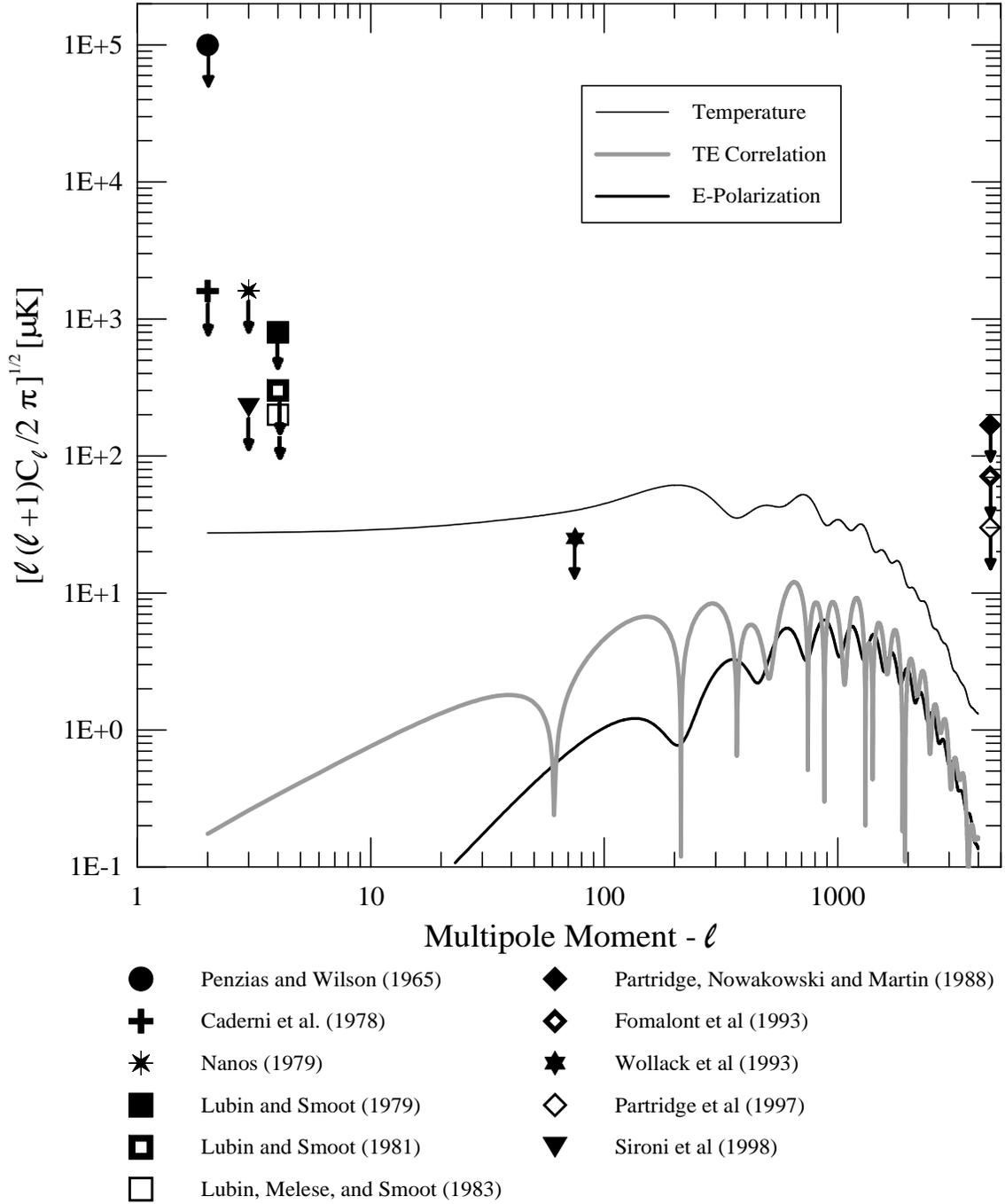}
\vspace{6.0in}
\caption{The state of the field:  approximate experimental limits on the CBR
polarization.  The same CDM model as in Figure \ref{fig-lvsnu}
is sketched in for reference.}\label{fig-polimit}
\end{figure}

\section{Overview of Techniques}

Figure \ref{fig-polimit} presents current experimental limits on the CBR
polarization.
The first experiments dedicated toward detecting or limiting the
CBR polarization (Caderni et al.\ 1978; Nanos 1979; Lubin \&
Smoot 1979, 1981;
Lubin, Melese \& Smoot 1983) modulated the polarization of the incoming
signal
(either with a Faraday switch or a mechanical rotating analyzer)
and locked in at the modulation frequency to measure $Q$ or $U\!$,
depending on the orientation of the polarimeter. The increased
sensitivity of bolometer-based and HEMT\footnote{HEMT stands for
high electron mobility transistor; amplifiers using HEMTs have
very low noise
(Pospieszalski 1992, 1995).}-amplifier-based
radiometers  being built today is such that the systematic
effects associated with the Faraday switch may be too
large to tolerate.  (Faraday switches are very sensitive to
temperature variations and external magnetic field variations.)
The modulation frequency of a mechanically rotating analyzer is
limited in practice.  Several new methods of Dicke-switching are
described below.

One other experimental limit appears in Figure \ref{fig-polimit}, the limit
from
the Saskatoon anisotropy experiment (Wollack et al.\ 1993),
at $\ell\approx 80$.  This
experiment collected temperature differences between patches on
the sky, $\Delta T_i$.  Measurements of $\Delta T_i$ were made in
two orthogonal linear polarizations, so that the polarization
anisotropy limit comes from comparing the two sets of $\Delta
T_i$.  Thus, the variance of the polarization anisotropy measured
includes a component due to the spatial variation of the
(unpolarized) atmosphere.  In other words, the best limit on the
plot comes from an experiment optimized to measure the
temperature anisotropy, not the polarization!\footnote{In fact,
later data from Saskatoon at a slightly larger $\ell$ give an
even lower limit to the polarization, but the full analysis has
not yet been completed. (Page 1999).}  The current crop
of polarization experiments (with the exception of some satellite
missions which do not have to contend with atmospheric spatial
variations) measure the difference between two orthogonal linear
polarizations within a single beam at a time.

Receivers used in the current experiments dedicated to
polarimetery of the CBR may be divided into two categories:
1) bolometer pairs read out in an AC bridge, such as the
bolometer-bridge polarimeter invented for use in the Polatron
experiment described below and 2)
HEMT-amplifier correlation receivers.  Correlation receivers
(Fujimoto, 1964;  Rohlfs \& Wilson, 1996)
comprise two types also: a) receivers in which the two signals
proportional to the electric fields
from the two orthogonal polarizations are multiplied together
directly in
a nonlinear device and b) receivers in which the two signals are
split and recombined with appropriate phase shifts into four
signals which are subsequently detected and differenced to give
$Q$ and $U$. The former type of correlation receiver is
typically less bulky and capable of larger bandwidths than the
latter, but requires care to ensure sufficient dynamic range.
All these receivers may be Dicke-switched rapidly.
In particular, the correlation receivers may be phase-switched at
several kilohertz, fast enough to overcome the $1/f$ knee of even
the highest frequency (90~GHz) HEMT amplifiers. The switching is
achieved by inserting a relative phase shift of $0\deg$ or
$180\deg$ between the two
incoming signals; the output of a correlation receiver is
proportional to $\cos\theta$, where $\theta$ is the total phase
shift between the two incoming signals.   Direct-multiplication
correlators are used in the POLAR and PIQU experiments, while the
Milano polarimeter and the SPOrt experiment use
detect-and-difference correlation schemes.  (Peter Timbie
pioneered the use of direct-multiplication correlation receivers
for CBR anisotropy in the 1980's:  reference Timbie.)

\section {Polarization Experiments at the Turn of the Millenium}

Due to the authors' familiarity with three of the current
ground-based polarization experiments, we devote unequal attention to those
in
what follows.  For completeness, we mention five other
experiments of which we are aware.  The frequency coverage and
$\ell$ coverage of these experiments is depicted in Figure
\ref{fig-lvsnu}.
Much of the other pertinent information is condensed in Table \ref{tbl-1}.
First we discuss ground-based experiments, beginning with the
experiment sensitive to the smallest angular scales.  Then we
discuss space-based experiments briefly.

\subsection{The VLA 8.4~GHz CBR Project.}

Partridge et al.\ (1997) recently completed imaging and analysis of a
40~arcmin$^2$ field at 8.44~GHz
with 6\arcsec resolution, using the VLA.
The signals were collected in circular polarization, and all four
Stokes parameters recovered. The data came from 159 hours of
observations; such a project is unlikely to be repeated.
This work follows up on
earlier work at lower resolution (Fomalont et al.\ 1993) and at
lower frequency (Partridge et al.\ 1988).
These three papers contain the only published CBR polarization
limits from interferometry to date; the results are summarized
in Figure \ref{fig-polimit}.

\subsection{The Polatron.}

The Polatron experiment is being built by a team comprising
members from Caltech,  Stanford University and Queen Mary and
Westfield College.  The Polatron will be used to  search for CBR
polarization on arcminute angular scales where the amplitude of
the  polarization power spectrum is expected to peak.
Observations will be made at a frequency  of 100\,GHz where
confusion from polarized foregrounds is expected to be a minimum.

The corrugated  entrance feed of the Polatron will be
located at the Cassegrain focus of the 5.5\,m dish at the Owens
Valley  Radio Observatory (OVRO), generating a $2.5'$ beam on the
sky.  A broad-band OMT  (orthomode transducer, see Chattopadhyay
et al.\  1999) splits the incoming signal into two  orthogonal
linear polarizations.  The backend of the polarimeter feeds two
bolometers, each  of which detects one of the linear polarization
states that comes from the OMT.  Metal-mesh  resonant filters
located between the OMT and the bolometers define a 20\% passband
centered at 96\,GHz.  An AC-bridge readout circuit (see, for
example, Holzapfel et al.\ 1997) will be used to difference the
outputs from the two bolometers.  At any instance, this
differencing scheme permits the measurement of a single Stokes
parameter (Q or U).  By  rotating the plane of polarization by
45$^\circ$ with a quartz half-wave plate located in  front of the
entrance feed, a second Stokes parameter (U or Q) is then
measured.

The silicon nitride metal-mesh bolometers (Bock et al.\ 1996) that
will be used in the  Polatron are identical to those that will be
flown as part of the Planck Surveyor High  Frequency Instrument
(HFI) in 2006.  The bolometers will be operated at 250\,mK where
they will have NEPs significantly lower than the background
photon noise limit. In 1  second of integration on a single
$2.5'$ pixel, the Polatron is expected to measure each  Stokes
parameter to a precision of 700\,$\mu$K.

The Polatron will be commissioned in the summer of 1999 and will
make its first  observations in the winter of 1999/2000.  In 6
months it is expected that the Polatron will  observe 850 $2.5'$
pixels to a sensitivity of 8\,$\mu$K per pixel in each of Q and
U.  In a  standard CDM model, this will be sufficient to detect
rms polarization at 5$\sigma$. Over its projected four-year
lifetime, the Polatron plans to map out the polarization anisotropy
between $\ell=200$ and $\ell=2600$.

\subsection{The PIQU Experiment.}
A group at Princeton University has designed an experiment
nicknamed PIQU, for Princeton IQU, a reference to the Stokes
parameters the experiment eventually plans to measure.
PIQU will measure the polarization at small angular scales at two
frequencies, 40~GHz and 90~GHz, with multiple horns in the focal
plane.
The experiment uses a 1.4~m off-axis parabola fed with a
corrugated horn antenna to provide a beamwidth of $0\fdg 23$
for the 90~GHz radiometer.  The RF signals from the two arms of
an OMT, corresponding to two linear orthogonal polarizations from
the sky, are mixed down to a 2--18~GHz IF, split into three
sub-bands,  and then directly
multiplied together in a broad bandwidth mixer. The front end of
the instrument uses HEMT amplifiers cooled to about 12~K.
Phase-switching at 4~kHz is done in one LO line.  Phase tuners in the IF
lines and the LO line allow balancing of the phase to $\pm
20\deg$, such that the bandwidth degradation is less than 10\%.

The phase-one
instrument, PIQ-90,  measures only one of $Q$ and $U$, chosen to be the
polarization with the most symmetry with respect to the
experimental apparatus, in an effort to minimize systematic
effects.  This instrument operates at 90~GHz and will be deployed
to the roof of Princeton in the spring of 1999. A separate 40~GHz
cryostat is being developed and will be operated in a follow-up
data run subsequently.  The 40~GHz receiver shares the IF components of
the 90~GHz receiver.  In subsequent years, multiple horns will be
placed in the focal plane, using a larger cryostat, so that
multiple multifrequency measurements may be made simultaneously.

\subsection {The POLAR Experiment.}\label{sec-polar}
The Polarizaton Observations of Large Angular Regions (POLAR)
experiment has been  designed and built by the Observational
Cosmology Group at the University of Wisconsin--Madison.  POLAR
is designed to measure the CBR's $Q$ and $U$ Stokes parameters in
several broad bands
between 26 and 100 GHz.  POLAR observes the CBR
polarization at large  ($7^{\circ}$ FWHM) angular
scales--comparable to the COBE satellite.  At these angular  scales the
rms CBR polarization signal is expected to be quite small ($<1
\mu$K) unless  the Universe was reionized at an early epoch
corresponding to a $z\sim100$.   The raw system
noise of  POLAR is a factor of 100 smaller than the instrument
that was used to set the current upper  limits on large angular
scales (Lubin, Melese \& Smoot, 1983).
The primary  goal of POLAR is to
reach a sensitivity level of  $\Delta$T$_{Pol}/$T$_{CBR}\le 10^{-
6}$.  At this level, POLAR will either detect polarization in the
CBR or place a tight  constraint on the epoch of reionization.

POLAR employs a correlation radiometer  that uses a corrugated feed horn to
couple the CBR into an orthomode transducer which  decomposes the
incoming radiation into two linear polarizations.  These two
linear  polarized components are amplified using
two HEMT amplifiers
cooled to $\sim$15 K in a cryocooler.
After  an additional stage of ambient temperature amplification,
the two parallel signal chains are  mixed down to an intermediate
frequency band (2-12 GHz) where they are multiplied.
The IF is subdivided into three equal bands for
additional foreground discrimination.  The output of the correlator
is proportional to $Q\sin 2\alpha +U\cos  2\alpha$ where $\alpha$
is the orientation of the polarimeter.  POLAR began observations  in the
Ka-band (26-36 GHz) in September 1998 at the University of
Wisconsin's Pine  Bluff Observatory.  POLAR observes $\sim 36$
spots directly overhead in a strip at a  declination of
$43\deg$.  Later in 1999, POLAR will add either a W-band (90-100
GHz)  or Q-band (35-45~GHz) polarimeter (in the same cryostat)
which will observe
simultaneously with the Ka-band  system.  This additional
polarimeter will help significantly in discriminating against
foreground contamination.  For additional information regarding
POLAR, see Keating et  al.\ 1998 and
{\it http://cmb.physics.wisc.edu/polar/.}  Future plans for the
POLAR team include collaborating with UC-Santa Barbara to develop
a small angular scale ($\approx 10'$) polarimeter.

\subsection{The Milano Polarimetry Project.}
Sironi et al.\ (1998) have built and operated a 33 GHz
polarimeter coupled to a $14\deg$ corrugated feed horn.  The
correlation receiver uses a phase discriminator comprising
several $90\deg$ hybrid tees (which output an incoming signal on
two ports which have a $90\deg$ phase difference between them)
and one $180\deg$ hybrid tee.  Both $Q$ and $U$ are observed
simultaneously.  An extension to the feed horn allows data to be
taken at $7\deg$; data have also been collected in this mode
(Sironi 1999).  Future plans include mounting the polarimeter at
the Cassegrain focus of the 2.6 m dish at Testa Grigia (Sironi et
al. 1998), which would allow measurements
with a $1\deg$ beam.

\subsection{Space-based Experiments.}

The MAP and Planck Surveyor satellites are designed to measure the CBR
temperature anisotropy across the entire sky.  Both will also
measure polarization.  MAP, which is set to launch in fall of
2000, should make a statistical detection of polarization
anisotropy.  Planck, with proposed launch date of 2007, will be
sensitive enough to detect polarization directly for most
CDM models.  SPOrt is an
experiment dedicated to measuring CBR polarization.  SPOrt is
meant to be deployed on the International Space Station in 2001 or 2002.

The MAP satellite (Jarosik et al. 1998) will use
polarization-sensitive radiometers to observe the whole sky, with a
frequency-dependent resolution ranging from $0\fdg21$ to
$0\fdg93$.  Data will be collected in five bands, from
22~GHz to 90~GHz.  The receivers are phase-switched at 2.5~kHz,
and use HEMT amplifiers.  Corrugated feed horns couple to OMTs to
separate two orthogonal linear polarizations.  The primary
reflectors are back to back $1.4\mbox{ m }\times1.6\mbox{ m}$
dishes.  The difference data consist of $T_A - T_{B^\prime}$ and
$T_B - T_{A^\prime}$ where $A$ and $A^\prime$ designate the two
orthogonal linear polarizations from horn A, and similarly for
$B$ and $B^\prime$.  From such pairs of numbers one can derive a
full-sky map of the polarization of the sky (Wright, Hinshaw, \&
Bennett 1995).  Most of the radiometers are completely
constructed and tested.  More information is available at
{\it  http://map.gsfc.nasa.gov/html/technical\_info.html}.

The Planck satellite (Tauber 1998) has two sets of receivers, the LFI set
and
the HFI set, for ``Low-Frequency Instrument" and ``High-Frequency
Instrument."  The HEMT-based LFI radiometers are all linearly polarized, and
include four bands between 30 and 100~GHz. The angular resolution
for the LFI ranges from $0\fdg55$ to $0\fdg20$.
The HFI receivers use bolometers.  Three of the six HFI
frequency bands are linearly polarized (143~GHz, 217~GHz and
545~GHz). The resolution for polarized bands ranges from
$0\fdg13$ to $0\fdg08$.
More information is available at
{\it http://astro.estec.esa.nl/SA-general/Projects/Planck/}.

The SPOrt project (Cortiglioni, et al. 1998) aims to measure the
polarization of over 80\% of the sky in $7\deg$ patches in
several frequency bands between 20 and 90~GHz.  The receivers are
correlation receivers of the detect-and-difference variety, using
phase-switching, and measuring $Q$ and $U$ simultaneously.

\section{Summary and Discussion}

In addition to the CBR brightness spectrum and the temperature
anisotropy, the CBR polarization represents the third treasure trove of
information encoded in the CBR. The small polarization signal raises
large barriers that undermine the ability to extract this information.
These barriers include obtaining the requisite statistical sensitivity,
minimizing systematic errors, and discriminating against foreground
contamination.  (See Tegmark, these proceedings.)

The experiments described above overcome these barriers in a variety of
different ways.
 Long integration times (roughly 10
hours/pixel/detector) with ever more sensitive detectors will yield the
statistical sensitivity needed to detect CBR polarization.
These experiments will have to minimize systematic effects at levels
that have yet to be charted.  The minimization of these systematic
effects impacts the design and implementation of experiments in many
different ways as demonstrated by the variety of experiments described
above.  The foregrounds that could contaminate CBR polarization
are not well characterized.
Polarized galactic synchrotron radiation is the only
known foreground, and its intensity (much less its polarization)  has
yet to be measured at high galactic latitudes at the relevant
frequencies.  The best way to marginalize this ignorance is to perform
multifrequency measurements, as temperature anisotropy experiments
have shown.
Since the {\em best} overall approach has yet to be determined, the
community will invariably benefit from the variety of
experimental approaches described above.

\begin{table}
\caption{Parameters of ongoing CBR polarization experiments are
compiled here. References are given in the text, except in those
cases where the information is communicated by
the authors.} \label{tbl-1}
\begin{center}\scriptsize
\begin{tabular}{lllll}
Experiment & beamsize\tablenotemark{a} & frequency &
Receiver\tablenotemark{b} & Site  \\
 & & GHz & &\\
\tableline
VLA & $0\fdg 02$ & 8.44  & int & NM desert \\
POLATRON & $0\fdg 04$ & 96  & bolo br & OVRO \\
PLANCK HFI & $0\fdg 08$ & 143, 217, 545 & bolo & space L2 \\
PIQU & $0\fdg 22$ & 40, 92  & HCdm & Princeton, NJ \\
PIQU2\tablenotemark{c} & $\la 0\fdg 2$ & 40, 92 & HCdm & high alt site \\
MAP  & $0\fdg 23$ & 22, 33, 40, 61, 98  & HTP  & space L2 \\
PLANCK LFI & $0\fdg 20$  & 30,44,70,100 & HTP & space L2 \\
POLAR & $7\deg$ & 30 \& 40 or 90  & HCdm & Madison, WI\\
SBUW\tablenotemark{c} & $\la 0\fdg 2$  & 30 \& 40 or 90 & HCdm & high alt US
\\
SPORT & $7\deg$ & 22, 32, 60, 90 & HCdd & space station \\
Milano & $7\deg, 14\deg$ & 33  & HCdd & Antarctica\\
Milano2\tablenotemark{c} & $1\deg$ & 33  & HCdd & Alps\\

\end{tabular}
\end{center}


\tablenotetext{a}{The smallest beamsize is given; most of the
experiments have frequency-dependent beamsizes.}
\tablenotetext{b}{Receiver types include {\bf int}
(inteferometer), {\bf bolo br} (bolometers in AC bridge),
{\bf bolo} (bolometers), {\bf HCdm} (HEMT correlation receivers with
direct multiplication), {\bf HCdd} (HEMT correlation receivers using
recombination of signals with appropriate phases, followed by
detection and differencing), and {\bf HTP} (HEMT total power -- see
text and the MAP website for more information.)}
\tablenotetext{c} {These three experiments are planned upgrades to
existing experiments; the exact parameters are not yet
determined.  SBUW is a collaboration between UC--Santa Barbara and the
existing POLAR team.}

\end{table}


\begin{references}

\reference Bock J.J., DelCastillo H.M., Turner A.D., Beeman J.W., Lange A.E.
\& Mauskopf P.D.,  1996, in proceedings of the 30th ESLAB
Symposium: {\em `Submillimetre and Far- Infrared Space
Instrumentation'}, SP-388, 119.

\reference  Caderni, N.  1978, Phys. Rev. D, 17, 1908.

\reference Chattopadhyay G., Philhour B., Carlstrom J., Church S., Lange A.
\& Zmudzinas J.,  1999, IEEE microwave and guided wave
letters, submitted.

\reference Cortiglioni, S. Cecchini, S., Carretti, E., Orsini, M., Fabbri,
R., Boella, G., Sironi, G., Monari, J., Orfei, A., Tascone, R.,
Pisani, U., Ng, K. W., Nicastro, L., Popa, L., Strukov, I. A.,
\& Sazhin, M. V. 1998, {\it astro-ph/9901362}.

\reference  Fomalont, E. B., Partridge, R. B. Lowenthal, J. D. \& Windhorst,
R. A. 1993, \apj , 404, 8.

\reference Fujimoto, K.  1964, IEEE-MTT, 203.

\reference Holzapfel W.L., Wilbanks T.M, Ade P.A.R., Church S.E., Fischer
M.L., Mauskopf  P.D., Osgood D.E. \& Lange A.E., 1997, \apj, 479,
17.

\reference Jarosik, N., Limon, M., Page, L., Spergel, D., Wilkinson, D.,
Bennett, C., Hinshaw, G., Kogut, A., Mather, J., Halpern, M.,
Meyer, S., Tucker, G., Wollack, E., \& Wright, E. 1998,
Proceedings of the XXXIIIrd Rencontres de Moriond, Editions
Frontiers, Paris, 249.

\reference Keating, B., Timbie, P., Polnarev, A., \&
Steinberger, J.  1998, \apj, 495, 580.

\reference  Lubin, P. M., \& Smoot, G. F.  1981, \apj, 245, 1.

\reference  Lubin, P. M., \& Smoot, G. F.  1979, \prl, 42, 2, 129.

\reference  Lubin, P., Melese, P. \& Smoot, G.  1983, \apj, 273, L51.

\reference  Nanos, G.  1979, \apj, 232, 341.

\reference Netterfield, C. B., Jarosik, N., Page, L., Wilkinson, D.,
\& Wollack, E.   1995, \apj,  445, L69.

\reference Page, L. A. 1999, private communication.

\reference  Partridge, R. B., Nawakowski, J., \& Martin, H. M. 1988, Nature,
311, 146.

\reference Partridge, R. B., Richards, E. A., Fomalont, E. B.,
Kellerman, K. I., \& Windhorst, R. A.  1997, \apj, 483, 38.

\reference  Penzias, A. \& Wilson, R.  1965, \apj , 142, 419.

\reference Pospieszalski, M.  1992, IEEE-MTT-S Digest, 1369.

\reference Pospieszalski, M. 1995, IEEE-MTT-S Digest, 1121.

\reference Rohlfs, K. \& Wilson, T. L.  1996, Tools of Radio
Astronomy, Springer-Verlag, Berlin.

\reference Seljak, Uro\v{s}, \& Zaldarriaga, Matias 1996, \apj,
469, 437. 

\reference Sironi, G., Boella, G. Bonelli, G., Brunetti, L., Cavaliere,
F., Fervasi, M., Giardino, G.,  \& Passerini, A.  1998, New Astronomy, 3, 1.

\reference Sironi, G. 1999, private communication.

\reference Tauber, J. A., 1998, Proceedings of the XXXIIIrd Rencontres de
Moriond, Editions Frontiers, Paris, 255.

\reference Wollack, E. J., Jarosik, N. C., Netterfield, C. B., Page, L. A.,
\& Wilkinson, D.  1993,  \apj, 419, L49.
\reference Wright, E. L., Hinshaw, G., \& Bennett, C. L. 1995, \apj, 458,
L53.

\reference Zaldarriaga, M., \& Seljak, U. 1997, \prd, 55, 1830.

\end{references}
\end{document}